\documentclass[10pt]{article}
\usepackage[cp866]{inputenc}
\begin{document}

\bigskip
\centerline {{\Large\bf Physical meaning and a duality of concepts of }} 
\centerline {{\Large\bf wave function, action functional, entropy, }} 
\centerline {{\Large\bf the Pointing vector, the Einstein tensor}} 
\centerline {\it L.~I. Petrova}
\centerline{{\it Moscow State University, Russia, e-mail: ptr@cs.msu.su}}

\renewcommand{\abstractname}{Abstract}
\begin{abstract}

Physical meaning and a duality of concepts of wave function,
action functional, entropy, the Pointing vector, the Einstein
tensor and so on can be disclosed by investigating the state of
material systems such as thermodynamic and gas dynamic systems,
systems of charged particles, cosmologic systems and others. These
concepts play a same role in mathematical physics. They are
quantities that specify a state of material systems and also
characteristics of physical fields. The duality of these concepts
reveals in the fact that they can at once be both functionals and
state functions or potentials. As functionals they are defined on
nonintegrable manifold (for example, on tangent one), and as a
state function they are defined on integrable manifold (for
example, on cotangent one). The transition from
functionals to state functions dicribes the mechanism of physical
structure origination. The properties of these concepts can be studied 
by the example of entropy and action. The role of these concepts in 
mathematical physics and field theory will be demonstrated.

Such results have been obtained by using skew-symmetric forms.
In addition to exterior forms, the skew-symmetric forms, which are obtained
from differential equations and, in distinction to exterior forms, are
evolutionary ones and are defined on nonintegrable manifolds, were used.

\end{abstract}

{\large {\bf Introduction}}

Such concepts as wave function, entropy, action, the Pointing vector,
the Einstein tensor and others are those on which the fundamental branches
of mathematical physics and field theories are based. In the present paper
we try to show that these concepts have a same physical meaning. 

The properties of conservation laws lie at the basis of the present 
investigation. Before beginning this investigation, it should be said 
a little about the concept of 'conservation laws'.

The concept of 'conservation laws' assumes a different meaning in
various branches of physics and mechanics.

In areas of physics related to the field theory and in theoretical
mechanics 'the conservation laws' are those according to which there
exist conservative physical quantities or objects. These are the
conservation laws that above were named 'exact'.

In mechanics and physics of continuous media the concept of 'conservation laws'
relates to the conservation laws for energy, linear momentum, angular momentum,
and mass that establish the balance between the change of physical quantities
and the external action. These conservation laws can be named 'the balance
conservation laws'. They are described by differential (or integral) equations.
It may be pointed out that all continuous media such as thermodynamic,
gas dynamical, cosmic systems and others (which can be referred to
as material systems), are subject to the balance conservation laws.

It turns out that the balance and exact conservation laws are
mutually connected. All physical processes are controlled by
interaction between the balance and exact conservation laws. The
concepts pointed above are functionals that describe the
controlling role of conservation laws in evolutionary processes
and in processes of generating physical fields. It appears to be
possible to disclose the physical meaning and duality of these
concepts by studying the relation that follows from the balance
conservation laws and describes the mechanism of transition from
balance conservation laws to exact ones.

The investigation has been carried out by using skew-symmetric
differential forms whose properties reflect the properties of
conservation laws. Exact conservation laws are described by closed
exterior skew-symmetric forms [1,2]. (It is known that the
differential of closed exterior form equals zero, that is, the
closed form is a conservative quantity). The Noether theorem,
which is expressed as $d\omega=0$, is an example. While studying
the balance conservation laws, the skew-symmetric differential
forms also arise. The relation that follows from the equations of
balance conservation laws is expressed in terms of skew-symmetric
forms. However, in contrast to exterior forms, these
skew-symmetric forms are defined on nonintegrable manifolds.

The physical meaning and duality of these concepts can be demonstrated while
analyzing the equations of balance conservation laws for energy and momentum.

\section{Properties of the balance conservation law equations}

In mechanics and physics of material systems the equations of balance
conservation laws are used for description of physical quantities,
which specify the behavior of material systems. However, the equations of
balance conservation laws not only describe the variation of physical
quantities. Their role in mathematical physics is much wider.

The required functions for the equations of balance conservation laws are
usually functions which relate to such physical quantities as the particle
velocity (of elements), temperature or energy, pressure and density [3-5]
Since these functions relate to one material system, it has to be a connection
between them. The concepts analyzed (entropy, action and others)
are quantities that describe this connection and characterize the system state.
The system state appears to be an equilibrium one if these concepts are functions
of state. The relation that follows from the balance conservation laws is just a
relation for such function of state. However, as it will be shown later, this
relation turns out to be nonidentical. And this means that the concepts described
are functionals (this corresponds to nonequilibrium system state). As it will be
shown, they become state functions only at realization of some conditions. And
to this case the degenerate transformation and the transition from nonintegrable
manifold to integrable one are assigned.

\subsection{Evolutionary nonidentical relation}

Let us analyze the equations that describe the balance conservation
laws for energy and linear momentum.

We introduce two frames of reference: the first is an inertial one
(this frame of reference is not connected with the material system), and
the second is an accompanying one (this system is connected with the manifold
made up by the trajectories of material system elements).

As one can see, for various material systems the energy equation
in the inertial frame of reference can be reduced to the form:
$$
\frac{D\psi}{Dt}=A \eqno(1)
$$
where $D/Dt$ is the total derivative with respect to time, $\psi $ is the
function of state that specifies the material system, $A$ is the quantity that
depends on specific features of the system and external energy actions on
the system.

It is well known that the equations for entropy and action have such a form.

So, the equation for energy presented in terms of the action  $S$ has a form:
$$
\frac{DS}{Dt}=\,L \eqno(2)
$$
Here $\psi \,=\,S$, $A\,=\,L$ where $L$ is the Lagrange function.

The equation for energy of an ideal gas can be presented in the form [5]:
$$
\frac{Ds}{Dt}=\,0 \eqno(3)
$$
where $s$ is the entropy. In this case $\psi \,=\,s$, $A\,=\,0$.

It is worth noting that the examples presented show that the action
functional and entropy play the common role.

At this point, the following should be noted. It is known that entropy can
be a function of the thermodynamic system state. In this case it depends on
thermodynamic variables. Besides of this, entropy can be a function of the state
of gas dynamic system. In this case it depends on the space-time coordinates.
The entropy in equation (3) is a characteristics of gas dynamical system.

In the accompanying frame of reference (connected with the manifold formed by
the trajectories of the  system elements) the total derivative
with respect to time is transformed into the derivative along the trajectory.
Equation (1) is now written in the form
$$
{{\partial \psi }\over {\partial \xi}}\,=\,A_{\xi}\eqno(4)
$$
here $\xi$ is the coordinate along the trajectory, $A_{\xi}$ is
equal $A$ in this case (see equation (1)). In the general case
$A_{\xi}$ is a quantity that depends on the energetic action to
the material system and on the system characteristics.

What is a peculiarity of the energy equation?

From this equation one can seemingly obtain a required state function.

However, it turns out that the required state function in addition has to
satisfy the equation of the conservation law for momentum, which in the
accompanying frame of reference is written as
$$
{{\partial \psi}\over {\partial \nu }}\,=\,A_{\nu },\quad \nu \,=\,2,\,...\eqno(5)
$$
where $\nu$ are the coordinates in the direction normal to the
trajectory, $A_{\nu }$ are the quantities that depend on the specific
features of the system and external force actions.

The function that simultaneously obeys both equations can be obtained
only in the case when these equations appear to be consistent. Since these
equations are for derivatives of the same function, they can be consistent
only if the derivatives made up a differential, i.e. the mixed derivatives
appear to be commutative.

Eqs. (4) and (5) can be convoluted into the relation for differentials
$$
d\psi\,=\,A_{\mu }\,d\xi ^{\mu },\quad (\mu\,=\xi,\nu )\eqno(6)
$$
where $d\psi $ is the differential expression
$d\psi\,=\,(\partial \psi /\partial \xi ^{\mu })d\xi ^{\mu }$.
Relation (6) can be written as
$$
d\psi \,=\,\omega \eqno(7)
$$
here $\omega \,=\,A_{\mu }\,d\xi ^{\mu }$ is the differential form of the
first degree.

This relation is an evolutionary one since it has been obtained from
the evolutionary equations.

The relation obtained possesses a peculiarity, namely, it turns out to be
nonidentical. The skew-symmetric form in the right-hand side of this relation
is not a closed form, i.e. a differential, like the left-hand side.
The differential form $\omega\,=\,A_{\mu }d\xi ^{\mu }$ appears to be unclosed
since the commutator of the form $\omega $, and hence a differential, are nonzero.
The components of the commutator of such a form can be written as follows:
$$
K_{\alpha \beta }\,=\,\left ({{\partial A_{\beta }}\over {\partial \xi ^{\alpha }}}\,-\,
{{\partial A_{\alpha }}\over {\partial \xi ^{\beta }}}\right )
$$
(here the term  connected with the manifold metric form
has not yet been taken into account).

The coefficients $A_{\mu }$ of the form $\omega $ have been obtained either
from the equation of the balance conservation law for energy or from that for
linear momentum. This means that in the first case the coefficients depend
on the energetic action and in the second case they depend on the force action.
In actual processes the energetic and force actions have different nature
and appear to be inconsistent. The commutator of the form $\omega $ made up by
the derivatives of such coefficients is nonzero.
This means that the differential of the form $\omega $ is nonzero as well.
Thus, the form $\omega$ proves to be unclosed and cannot be a differential
like the left-hand side of relation (7). The evolutionary relation (7) appears
to be nonidentical: the left-hand side of this relation is a differential,
whereas the right-hand side is not a differential.
[Nonidentity of such relation has been pointed out in the monograph [6]. In
that case a possibility of using a symbol of differential in the left-hand
side of this relation is allowed.]

The nonidentity of the evolutionary relation points out to the fact that
the state function differential does not exist. From this it follows that the
function desired is not a state function but is a functional, which depends
on the nonzero commutator of unclosed form in the right-hand side of the 
evolutionary relation. The functional depends on the quantities that make 
a contribution to the commutator such as the energetic or force actions on 
the system. (If the commutator be zero, the form  $\omega $ would be closed 
and be a differential. In this case one would be able to define 
the differential $d\psi$ and obtain the state function $\psi$.)

Thus we have that the quantity, which characterizes the material
system state and for which entropy and action are the examples,
turns out to be a functional. This points out to the fact that the
material system state is a nonequilibrium one. (The nonidentity of
the evolutionary relation obtained from the equations of balance
conservation laws means that the equations of balance conservation
laws are inconsistent. This points out to a noncommutativity of
the conservation laws. It is a noncommutativity of the
conservation laws that is just a cause of the nonequilibrium state
of material systems.)

The evolutionary relation possesses one more peculiarity,
namely, this relation is a selfvarying relation. (The evolutionary form
entering into this relation is defined on the deforming manifold
made up by trajectories of material system elements. This means
that the evolutionary form basis varies. In turn,
this leads to variation of evolutionary form, and the process
of intervariation of the evolutionary form and the basis is
repeated.) It should be emphasized that under selfvariation the evolutionary
relation remains to be nonidentical since the evolutionary form in the
right-hand side of the relation remains to be unclosed.

The selfvariation of evolutionary relation points out to the fact that
the required function, which characterizes the material system state,
changes but remains to be a functional. This means that
the material system state changes but keeps to be nonequilibrium.

From the properties of nonidentical relation it follows that under
degenerate transformation the identical relation can be obtained
from that. The degrees of freedom of material systems (translational,
rotating, oscillating and others) are the conditions of degenerate
transformation. The conditions of degenerate transformation specify the
pseudostructures (the integral surfaces): the characteristics (the determinant
of coefficients at the normal derivatives vanishes), the singular
points (Jacobian is equal to zero), the envelopes of
characteristics of the Euler equations and so on. (Sections of the cotangent
bundles, cohomologies by de Rham, singular cohomologies, potential surfaces, 
eikonals are examples of pseudostructures and relevant surfaces.)

The realization of the conditions of degenerate transformation leads to 
realization of some pseudostructure $\pi$ (the closed dual form) and formatting the
closed inexact form $\omega_\pi$. On the pseudostructure $\pi$ from
evolutionary relation (7) it is obtained the relation
$$
d\psi_\pi=\omega_\pi\eqno(8)
$$
which turns out to be identical since the form in the right-hand side
is closed and hence is a differential.
(It should be noted that such a differential is an interior one: it asserts
only on pseudostructure, which is defined by the condition of degenerate
transformation). Since the relation (8) is an identical one, from that one can
obtain the differential $d\psi_\pi$, and this points to the availability of
the state function desired.

Under realization additional conditions, which may be caused by the degrees
of freedom of material system, the identical on pseudostructure relation
is obtained from the nonidentical evolutionary relation. In this case the original
relation remains to be nonidentical. At this point it should be noted that
the original relation is defined on the manifold made up by the trajectories
of material system elements, whereas the identical relation is defined on
pseudostructure. The manifold made up by the trajectories of material system
elements is a deforming nonintegrable manifold, and the pseudostructures
form an integrable manifold. That is, the degenerate transformation is a transition 
from nonintegrable manifold to integrable one. On pseudostructure the function
desired (entropy and action are examples), which characterizes the system state, 
appears to be a state function (or a potential). But in this case remains to 
be a functional on the original manifold.

The availability on pseudostructure of the state function points out to
a locally-equilibrium state of the system, whereas the total state of
material system remains to be nonequilibrium.

One can see that the transition from the functional to state function
describes the process of realization of the locally-equilibrium state of the system.

Such a transition is accompanied by the origination in material system of
any observable formations such as waves, vortices, turbulent pulsations and so on.
The intensity of these formations is defined by the commutator of unclosed evolutionary
form.

This follows from the analysis of the identical relation (8).

Identical relation (8) holds the duality. The left-hand side of this
relation includes the differential, which specifies material system
and whose availability points to the locally-equilibrium state of
material system. And the right-hand side includes a closed inexact
form, which is a characteristics of physical structures. Physical fields are
formatted of such physical structures. (Closed inexact form and relevant closed
dual form describe a quantity being conservative on pseudostructure. Physical
structures that create physical fields are pseudostructures with a
conservative physical quantity.)

Thus we have obtained that the transition from the nonequilibrium state to the
locally equilibrium one is accompanied by emergence of
physical structure, which reveals in material system as an emergence of
certain observable formations (which develop spontaneously).
(Massless particles, structures made up by eikonal surfaces and wave fronts,
and so on are examples of physical structures.)

[The observed formation and the physical structure are
not identical  objects. If the wave be such a formation, the element of wave
front made up the physical structure at its motion.

Structures of physical fields and the formations of material systems
observed are a manifestation of the same phenomena. The light
is an example of such a duality. The light manifests itself
in the form of a massless particle (photon) and as a wave.]

Thus, one can see that the nonidentical evolutionary relation for functionals
and the transition from functionals to state functions disclose the mechanism
of emergence of physical structures, which made up physical fields, and
the process of emergence of various observable formations in material media.

\bigskip
Relation (8) has been obtained from the equations of balance conservation
law for energy and linear momentum. In this relation the form $\omega $
is that of the first degree. If the equations of the balance conservation
law for angular momentum be added to the equations for energy and linear
momentum, this form  will be a form of the
second degree. And, in combination with the equation of the balance
conservation law for mass, this form will be a form of degree 3.
In general case the evolutionary relation can be written as
$$
d\psi \,=\,\omega^p \eqno(9)
$$
where the form degree  $p$ takes the values $p\,=\,0,1,2,3$.
{\footnotesize (The relation for $p\,=\,0$ is an analog to
that in the differential forms, and it has been obtained from the
interaction of energy and time.)}

The relation (9), as well as the relation (8), is nonidentical.
This relation discloses the properties of such concepts as wave function
($p=0$), the Pointing vector ($p=2$), the Christoffel and Einstein tensors
($p=3$).

\bigskip

Below we analyze some peculiarities of entropy and action.

\section{Peculiarities of entropy}

In the next subsections the properties of entropy as a functional and a state function
of thermodynamic system and gas dynamic system are described.

\subsection{Entropy as a functional and a state function of thermodynamic
system}

It is known that the first and second principles of thermodynamics, which have been
introduced as postulates, form the basis of thermodynamics[7].

The first principle of thermodynamics is an example of the evolutionary
nonidentical relation. It follows from the balance conservation laws for
energy and linear momentum and is valid in the case when the heat influx
is the only external action. The second principle of thermodynamics with
equality is an example of identical relation. This relation follows from
the first principle of thermodynamics when the condition of integrability,
i.e. the realization of the integrating multiplier, namely, temperature, is
satisfied. The condition of integrability is a condition of degenerate
transformation at which the entropy as state function is realized. The second
principle of thermodynamics with inequality takes into account the availability
in real processes another actions on the system in addition to the heat influx
and is an example of nonidentical relation. In this case entropy is a functional.

\bigskip

Is is well known that the first principle of thermodynamics
can be written in the form
$$dE+\delta w\,=\,\delta Q $$
where $dE$ is the variation of the thermodynamic system energy, $\delta w$
is the work made by the system (this means that $\delta w$ is expressed in terms
of the system parameters), $\delta Q$ is the heat delivered to the system
(i.e. the external action). Since the term $\delta w$ must be expressed in terms
the system parameters and characterizes a real (rather then virtual) variation,
it can be denoted as $dw$, the first principle of thermodynamics
can be written as
$$dE\,+\,dw\,=\,\delta Q\eqno(10)$$

What is a distinction of the first principle of thermodynamics from the
conservation laws?

The balance conservation law for the energy of thermodynamic system can be
written in the form
$$ dE\,=\,\delta Q\,+\,\delta G \eqno(11)$$
where by $G$ we denote others (in addition to the heat influx) power actions.
For thermodynamic system the balance conservation law for linear momentum
(the variation of the system momentum as a function of force and mechanical
actions on the system) can be written as

$$dw\,=\,\delta W \eqno(12)$$
Here the force (mechanical) action on the system (external compression
of the system, effect of boundaries and others for example) is denoted
by $\delta W$.

If to sum relations (11) and (12), we obtain the relation
$$dE\,+\,dw\,=\,\delta Q\,+\,\delta G\,+\,\delta W \eqno(13)$$
which is just the evolutionary relation for thermodynamic system.

By comparing the relation (13), which follows from the
balance conservation laws for energy and linear momentum, with the relation (10),
one can see that they are identic if the heat influx is the only external action
on thermodynamic system ($\delta W=0$ and $\delta G \,=\,0$).

Thus, the first principle of thermodynamics follows from the balance
conservation laws (and not only corresponds to the conservation law for energy).
This is analogous to the evolutionary relation.

Since $\delta Q$ is not a differential (closed form), the relation
(10), which corresponds to the first principle of thermodynamics,
as well as the evolutionary relation, appears to be a nonidentical
nonintegrable relation. The form $dE\,+p\,dV$, even it is made up
of differentials, in the general case without the integrating
multiplier, is not a differential, since its terms depend on
different variables, namely, the first term is defined by
variables that specify the internal structure of the element, and
the second term is defined by the variables, which characterize
the interaction of elements, for example, such as the pressure.

In the this case under consideration entropy doesn't explicitly enter in relation (10).
In this case the state of thermodynamic system is defined by the expression
$dE\,+p\,dV$, which is not a differential, and this points out that there is no
the state function.

As it follows from the analysis of evolutionary relation, the
state function can be obtained only under degenerate
transformation. To this it has to correspond the realization of
the additional condition.

\bigskip

Consider the case when the work performed by system proceeds
through a compression. Then $dw\,=\,p\,dV$ (here $p$ and $V$ are
the pressure and the volume) and $dE\,+\,dw\,=\,dE\,+\,p\,dV$. As
it is well known, the form $dE\,+p\,dV$ can become a differential
(a closed form) only if there is an integrating multiplier
$1/\theta\,=\,pV/R$, where $T$ is a quantity that depends only on
the system characteristics and is named the thermodynamic
temperature [7]. In this case the form $(dE\,+\,p\,dV)/T$ proves
to be a differential (interior) of a certain quantity. Such a
quantity is the entropy $S$:
$$(dE\,+\,p\,dV)/T\,=\,dS \eqno(14)$$

If the integrating multiplier $1/\theta=T$ is realized (this is
just a condition of degenerate transformation), that is, the
relation (14) is satisfied, from the relation (10), which
corresponds to the first principle of thermodynamics, it follows
that
$$dS\,=\,\delta Q/T \eqno(15)$$

This is just the second principle of thermodynamics for reversible processes.
This takes place only when the heat influx is the only action on the system.

If, in addition to the heat influx, the system experiences any mechanical
actions $\delta W$ (for example, the influence of boundaries) or the additional
power action $\delta G$, according to relation (13) from relation (14) it
follows
$$dS\,=\,(dE+p\,dV)/T\,=\,(\delta Q+\delta W+\delta G)/T $$
from which one finds
$$dS\, >\,\delta Q/T \eqno (16)$$
that corresponds to the second principle of thermodynamics for irreversible
processes.

{\footnotesize (It should keep in mind that in relation (14) the
differential is incomplete. This is valid only at availability of
integrating multiplier, namely, the inverse temperature, which is
expressed in terms of the system characteristics.)}

In the case analyzed, the differential of entropy, rather then the entropy itself,
becomes a closed form.
{\footnotesize \{In this case the entropy reveals as the thermodynamic potential, i.e. the state
function. To the pseudostructure it is assign the equation of state, which
defines the temperature dependence on thermodynamic variables\}}.

For the entropy becomes a closed form, one more condition has to be realized.
As such a condition it can be the realization of integrating direction, the
example of which is the sound speed: $a^2\,=\,\partial p/\partial \rho\,=\,\gamma\,p/\rho$.
In this case it is fulfilled the equality
$ds\,=\,d(p/\rho ^{\lambda })\,=\,0$, from which it follows that the entropy
$s\,=\,p/\rho ^{\lambda }\,=\,const$ itself is a closed form (of zero degree).

{\footnotesize \{However, this does not mean that the gas state is
isoentropic identically. The entropy is constant only along the
integrating direction (for example, in a adiabatic curve or on the
front of a sound wave), whereas along the direction being normal
to the integrating direction the normal derivative of entropy have
a discontinuity.\}}.

The transition from variables $E$ and $V$ to variables $p$ and $\rho$ at
the realization of integrating direction is a degenerate transformation.

It should be noted that both temperature and the sound speed are
not a continuous thermodynamic variables. The are variables that
are realized in thermodynamic processes when the thermodynamic
system possesses the degrees of freedom. One can see a similarity
of the temperature and the sound velocity, namely, the temperature
is an integrating multiplier and the sound velocity is an
integrating direction. \{It should be emphasized that in real
processes the total state of thermodynamic systems is a
nonequilibrium one, that is, the commutator of the form
$dE\,+p\,dV$ is nonzero. A quantity, which is described by the
commutator and acts as internal force, can increase. Prigigine [8]
defined this as 'the production of the excess entropy'. Just such
an increase of internal force is perceived as an increase of
entropy in irreversible processes.

The closed static system without external action can tend to the
state of total thermodynamic equilibrium. To this process it
corresponds the tendency of the state functional to its asymptotic
maximum. For dynamic system the tendency of the system to the
state of total thermodynamic equilibrium can be violated by the
dynamic processes and transitions to the state of local
equilibrium.\}

\subsection{Entropy as a functional and a state function of gas dynamic system}

Above we analyzed the peculiarities of entropy as a functional and a state
function of thermodynamic system. Such entropy depends of the thermodynamic variables.
Entropy is also a functional and a state function of gas dynamic system.
However, in this case it depends on the space-time coordinates.

Equations and the relation for entropy of gas dynamic system have been investigated and
presented in the paper [9]. Below we will consider the simplest case, namely,
a flow of ideal (inviscous, heat nonconductive) gas.

Assume that the gas is a thermodynamic system in the state of local equilibrium
(whenever the gas dynamic system itself may be in nonequilibrium
state), that is, the following relation is fulfilled [7]:
$$Tds\,=\,de\,+\,pdV \eqno(17)$$
where $T$, $p$ and $V$ are the temperature, the pressure and the gas volume,
$s$ and $e$ are entropy\index{entropy} and internal energy per unit volume.
The entropy $s$ in relation (17) is a thermodynamic
state function and depends on the thermodynamic variables.
For the gas dynamical system the thermodynamic state function
describes only the state of the gas dynamical element (a gas particle).
As it has been noted before and will be shown below, for the gas dynamical
system the entropy is also a state function, . However, in  this case the entropy is
a function of space-time coordinates.

The equation of conservation law for energy of ideal gas can
be written as
$${{Dh}\over {Dt}}- {1\over {\rho }}{{Dp}\over {Dt}}\,=\,0 \eqno(18)$$
where $D/Dt$ is the total derivative with respect to time (if to designate
the spatial coordinates by $x_i$ and the velocity components by $u_i$,
$D/Dt\,=\,\partial /\partial t+u_i\partial /\partial x_i$). Here  $\rho=1/V $
and $h$ are respectively the mass and enthalpy densities of the gas.

Expressing enthalpy in terms of internal energy $e$ with the help of formula
$h\,=\,e\,+\,p/\rho $ and using relation (17), the balance conservation law
equation can be put to the form
$${{Ds}\over {Dt}}\,=\,0 \eqno(19)$$ 

Since the total derivative with respect to time is that along the trajectory,
in the accompanying frame of reference (that is connected with the manifold
made up by the trajectories of the  system elements) equations (19)
take the form:
$${{\partial s}\over {\partial \xi}}\,=\,A_{\xi} \eqno (20)$$
where $\xi$ is the coordinate along the trajectory, $A_{\xi}=0$ (see
equation (19)).

In the accompanying frame of reference the equation of the conservation law
for linear momentum can be presented as [5]
$${{\partial s}\over {\partial \nu}}\,=\,A_{\nu }\eqno(21)$$
where $\nu$ is the coordinate in the direction normal to the trajectory, and
in two-dimensional case $A_{\nu}$ has the form [5,10]:
$$A_{\nu}={{\partial h_{0}}\over {\partial \nu}}+{({u_1}^2+{u_2}^2)}^{1/2}\zeta-
F_{\nu}+{{\partial U_{\nu}}\over {\partial t}} \eqno (22)$$
where $\zeta={{\partial u_{2}}\over {\partial x}}-{{\partial u_{1}}\over {\partial y}}$.  

Equations (20) and (21) can be convoluted into the relation
$$ds\,=\,\omega \eqno(23)$$
where $\omega\,=\,A_{\xi}d\xi+A_{\nu}d\nu$ is the first degree differential
form.

Relation (23) is an example of the nonidentical evolutionary relation (7).
While describing actual processes relation (23) turns out to be not identical,
since the evolutionary form $\omega $ is not closed and is not a differential,
its commutator is nonzero. From the analysis of
the expression $A_{\nu }$ and with taking into account that $A_{\xi}\,=\,0$
one can see that the terms related to the multiple connectedness
of the flow domain (the second term of expression (22)), the nonpotentiality
of the external forces
(the third term in (22)) and the nonstationarity of the flow (the forth term
in (22)) contribute to the commutator. All these factors lead to the emergence of
internal forces, the nonequilibrium state and developing the instability.

The nonidentity of this relation points to the fact that the entropy is a functional
since the differential of entropy does not exist.

From the properties of nonidentical relation it follows that under
degenerate transformation the identical relation (from which the differential of entropy
is defined) can be obtained
from that. The degrees of freedom of gas dynamical systems (translational,
rotating, oscillating and others) are the conditions of degenerate
transformation. The conditions of degenerate transformation specify
the integral surfaces (pseudostructures): the characteristics (the determinant
of coefficients at the normal derivatives vanishes), the singular
points (Jacobian is equal to zero), the envelopes of characteristics
of the Euler equations and so on. As it was already noted, the realization of
the conditions of degenerate transformation leads to realization of
pseudostructure $\pi$ (the closed dual form) and formatting the
closed inexact form $\omega_\pi$. On the pseudostructure $\pi$ from
evolutionary relation (23) it is obtained the identical relation
$$
ds_\pi=\omega_\pi\eqno(24)
$$
from which the differential $ds_\pi$ can be obtained. This means that
we have the realization of entropy as a state function of gas dynamic system,
whose availability points to the locally-equilibrium state of the gas
dynamic system. [In the text-books on gas dynamics it is assumed that from
equation (20) one can obtain the entropy along the
trajectory. However, it occurs that the entropy as a function
of space-time coordinates (that is, as a function of
the state) has in addition to satisfy equation (21)].

The realization of gas dynamic state function (entropy as a function
of space-time coordinates) points out to the transition from the nonequilibrium
state to the locally equilibrium one. This process is accompanied by emergence
of the gas dynamic observable formations such as waves, vortices and so on.
In this case the quantity, which is described by the commutator of unclosed form
$\omega$ and acts as an internal force (producing the nonequilibrium system state),
defines the intensity of these formations.

One can see that in gas dynamical system, even in the case of ideal gas,
it is possible the origination of formations that lead to emergence of vorticity.

In the case of viscous gas, the evolutionary form commutator will
contain the terms related to viscosity and heat conduction.
These terms are responsible for emergence of turbulent pulsations.

[Studying the instability on the basis of the analysis of entropy
behavior was carried out in publications by Prigogine and co-authors [8,11].
In their papers the entropy was considered as a thermodynamic function
of state (though its behavior along the trajectory was analyzed).
By means of such state function one can trace the development (in gas
fluxes) of the thermodynamic instability only [8]. To investigate the gas
dynamic instability it is necessary to consider the entropy as a gas dynamic
state function, i.e. as a function of the space-time coordinates.]

\section{Peculiarities of 'action'}

As it has been already noted, equation (2) for the action $S$ and
equation (3) for entropy of gas dynamical system obtained from the
balance conservation law for energy have the identic form.

The action, as well as the entropy, can be both a functional and a state
function.

In classic mechanics the action may be written in two forms: in
the Lagrangian form
$$S\,=\,\int\,L(q_j,\,\dot q_j)\,dt \eqno(25)$$
and in the Hamiltonian one
$$S\,=\,\int\,H(t,\,q_j,\,p_j)\,dt \eqno(26)$$
where $L$ is the Lagrangian function and $H$ is the Hamiltonian function
($H(t,\,q_j,\,p_j)\,=\,p_j\,\dot q_j-L$, $p_j\,=\,\partial
L/\partial \dot q_j$.

It is assumed that both forms are equivalent. However, these forms
are not equivalent. In expression (25) the action is defined on the Lagrangian
manifold $q_j,\, \dot q_j$, which is a tangent nonintegrable manifold. 
The action defined on such a manifold is a functional. In expression (26) 
the action is defined on the Hamiltonian
manifold $q_j,p_j$, which is a cotangent integrable manifold. 
The action defined on such a manifold is a function of state.

Here the degenerate transformation (the transition from the functional to the state
function)
is a transition from the Lagrangian function to the Hamiltonian function.
The transition from the Lagrangian function $L$ to the Hamiltonian function $H$
(the transition from variables $q_j,\,\dot q_j$ to variables
$q_j,\,p_j=\partial L/\partial \dot q_j$) is a transition from the tangent
manifold to the cotangent one.  This is a Legendre transformation, which is
a degenerate transformation.

In the equations of field theory
$${{\partial s}\over {\partial t}}+H \left(t,\,q_j,\,{{\partial s}\over {\partial q_j}}
\right )\,=\,0,\quad
{{\partial s}\over {\partial q_j}}\,=\,p_j\eqno(27)$$
the field function $s$ is an action, which is a function of state and is obtained from
the action functional $S\,=\,\int\,L\,dt$.
To equation (27) it is assigned the differential (a closed form)
$ds\,=\,-H\,dt\,+\,p_j\,dq_j$ (the Poincare invariant). This points out to
that the action $S$, which obeys the equation (26), is a function of state.

In quantum mechanics (where to the coordinates $q_j$, $p_j$ the operators are
assigned) the Schr\H{o}dinger equation for wave function serves as
an analog to equation (27). Wave function in quantum mechanics plays the same role
as the action in field theory.

\bigskip

The duality and physical meaning of concepts of entropy and action
were described above. It has been shown that these concepts relate
to the properties of the balance conservation laws for energy and
momentum, which are described by the evolutionary nonidentical
relation (7). As it was already noted, an account for another
balance conservation laws (of angular momentum and mass) leads to
relation (9), which is also evolutionary and nonidentical. The
relation (9) with skew-symmetric form $\omega $ of appropriate
degree $p$ discloses the properties of such concepts as wave
function ($p=0$), the Pointing vector ($p=2$), Christoffel's and
Einstein's tensors ($p=3$). All these concepts possess a duality.
Due to this duality all these concepts play a fundamental role in
the description of evolutionary processes in material systems and
the processes of generation of physical fields.

\section{A role of functionals in mathematical physics and field theory}

The functionals such as  wave function, action functional, entropy, the Pointing vector
and others are concepts that enable one to describe the regulating role of the
conservation laws in the evolutionary processes in material media and the
processes of generation of physical fields. The peculiarity of such functional
consists in that they are used both in the theories that describe material
systems (in mechanics and physics of continuous media) and in field theory.
As it is seen from the analysis of the evolutionary nonidentical relation
obtained from the equations of balance conservation laws for material systems,
in the theories that describe material systems these functionals specify
the state of material system. And in the field theory they describe physical
fields. One can see that the equations
of field theory are those in these functionals.

This duality of functionals just allows
to disclose a connection between the equations for material systems and
the field theory equations, which describe physical fields.
And this connection is realized with the help of nonidentical evolutionary relations.

The peculiarity of the field theory equations consists in the fact that all
these equations have the form of relations. They can be relations in
differential forms or in the forms of their tensor or differential analogs
(i.e. expressed in terms of derivatives).

The Einstein equation is a relation in differential forms. This equation
relates the differential of the form of first degree (Einstein's tensor)
and the differential form of second degree, namely, the energy-momentum
tensor. (It should be noted that Einstein's equation is obtained from
differential forms of third degree).

The Dirac equation relates  Dirac's \emph{ bra-} and \emph{ cket}- vectors,
which made up the differential forms of zero degree.

The Maxwell equations have the form of tensor relations.

Field equation and Schr\H{o}dinger's equation have the form of relations
expressed in terms of derivatives and their analogs.

All equations of existing field theories have the form of nonidentical
relations for the above pointed functional. From these equations
the identical relations follow, and from these relations
the differentials of functionals and closed exterior forms assigned to
the exact conservation laws for physical fields are obtained.
The identical relations that include  closed exterior forms or their
tensor or differential analogs are the solutions to the field theory equations.

As one can see, from the field theory equations it follows such identical
relation as

1) The Dirac relations made up of Dirac's \emph{ bra-} and
\emph{ cket}- vectors, which connect a closed exterior form of zero degree;

2) The Poincare invariant, which connects a closed exterior form of
first degree;

3) The relations $d\theta^2=0$, $d^*\theta^2=0$ for closed exterior
forms of second degree obtained from the Maxwell equations.

From the Einstein equation the identical relation is obtained in
the case when the covariant derivative of the energy-momentum
tensor vanishes.

It turns out that all equations of existing field theories are, in
essence, relations that connect skew-symmetric forms or their
analogs. In this case one can see that the equations of field
theories have the form of relations for functionals such as wave
function (the relation corresponding to differential form of zero
degree), action functional (the relation corresponding to
differential form of first degree), the Pointing vector (the
relation corresponding to differential form of second degree).
The tensor functionals that correspond to Einstein's
equation are obtained from the relation connecting the
differential forms of third degree.

The nonidentical evolutionary relation derived from the equations for
material media unites the relations for all these functionals. That is,
all equations of field theories are an analog of the nonidentical
evolutionary relation obtained from the equations of balance conservation laws.
From this it follows that the nonidentical evolutionary relation can
play a role of the equation of general field theory that discloses
common properties and peculiarities of existing equations of field
theory.

The correspondence between the equations of existing field
theories and the nonidentical evolutionary relations for the
functionals under consideration has the mathematical and physical
meaning. Firstly, this discloses the internal connection between
all physical theories, and, secondly, this enables one to
understand the basic principles of field theories, namely, their
connections with the equations for material systems generating
relevant physical fields. 

\bigskip

\bigskip

\centerline{\large\bf{Conclusion}}

The duality of such concepts as wave function, action functional, entropy,
the Pointing vector and others makes itself evident in the fact that, firstly,
they can at once be both functionals and state functions (or potentials)
and, secondly, as functionals they are used both in the theories that describe material systems (in mechanics and physics
of continuous medium) and in field theory.

Such a duality leads to that they serve not only for the description of
material media and physical fields but they also disclose the mechanism
of origination of various discrete structures and formations, which connect
physical fields with material media. The process of transition from functionals to
state functions (or potentials) describes the mechanism of origination of various structures.

Here the following should be emphasized.

Only due to the properties of skew-symmetric forms
defined on nonintegrable manifolds it turns out to be possible to disclose
the mechanism of origination of various structures described by the process
of transition from functionals to state functions (and connected with the
degenerate transformation and the transition from nonintegrable manifold
to integrable one).

Such skew-symmetric forms, which are evolutionary ones, are obtained from
differential equations and possess the unique properties. The mathematical
apparatus that is based on the properties of evolutionary and exterior skew-symmetric
form includes nontraditional elements such as nonidentical relations, degenerate
transformations and the transition from nonintegrable manifolds to integrable ones.
Such elements, which are contained in non of existing mathematical formalisms, enables one to
describe the discrete transitions, the origination of structures, the transitions
from nonconjugated operators to conjugated ones and others.

1. Cartan E., Les Systemes Differentials Exterieus ef Leurs Application
Geometriques. -Paris, Hermann, 1945.

2. Schutz B.~F., Geometrical Methods of Mathematical Physics. Cambridge
University Press, Cambrige, 1982.

3. Tolman R.~C., Relativity, Thermodynamics and Cosmology. Clarendon Press,
Oxford,  UK, 1969.

4. Fock V.~A., Theory of space, time and gravitation. -Moscow,
Tech.~Theor.~Lit., 1955 (in Russian).

5. Clark J.~F., Machesney ~M., The Dynamics of Real Gases. Butterworths,
London, 1964.

6. Synge J.~L. Tensorial Methods in Dynamics. -Department of
Applied Mathematics, University of Toronto, 1936.

7. Haywood R.~W., Equilibrium Thermodynamics. Wiley Inc. 1980.

8. Prigogine I., Introduction to Thermodynamics of Irreversible
Processes. -C.Thomas, Springfild, 1955.

9. Petrova L.I., The mechanism of generation of physical structures.
// Nonlinear Acoustics - Fundamentals and Applications
(18th International Symposium on Nonlinear Acoustics, Stockholm, Sweden,
2008) - New York, American Institute of Physics (AIP), 2008, pp.151-154.

10. Liepman H.~W. and Roshko ~A., Elements of Gas Dynamics. Jonn Wiley,
New York, 1957.

11.  Glansdorff P. and Prigogine I. Thermodynamic Theory of Structure,
Stability and Fluctuations. Wiley, N.Y., 1971.

\end{document}